# CONCEPTUAL DESIGN OF FIRST TOROIDAL ELECTRON CYCLOTRON RESONANCE ION SOURCE AND MODELING OF ION EXTRACTION FROM IT

C. Caliri[1,*], D. Mascali[2] and F.A. Volpe[1,†]

[1] *Department of Applied Physics and Applied mathematics, Columbia University, New York, NY, 10027 USA*

[2] *Istituto Nazionale di Fisica Nucleare, Laboratori Nazionali del Sud, 95123 Catania, Italy*

**Abstract**

Electron Cyclotron Resonance Ion Sources (ECRIS) progressed to higher and higher ion currents and charge states by adopting stronger magnetic fields (beneficial for confinement) and proportionally higher ECR frequencies. Further improvements would require the attainment of "triple products" of density, temperature and confinement time comparable with major fusion experiments. For this, we propose a new, toroidal rather than linear, ECRIS geometry, which would at the same time improve confinement and make better use of the magnetic field. Ion extraction is more complicated than from a linear device, but feasible, as our modelling suggests: single-particle tracings showed successful extraction by at least two techniques, making use respectively of a magnetic extractor and of **E** × **B** drifts. Additional techniques are briefly discussed.

[*] Present address: Istituto Nazionale di Fisica Nucleare, Laboratori Nazionali del Sud, 95123 Catania, Italy
[†] Corresponding author, fvolpe@columbia.edu

# 1. Introduction

Electron Cyclotron Resonance Ion Sources (ECRIS) are well-known electron-cyclotron-heated plasma-based ion-sources for accelerators and various applications ranging from nuclear fusion to industrial ion processing, isotope separation and mass spectroscopy [1]. The typical ECRIS device (Fig.1a) confines the plasma by means of a magnetic mirror and multipolar (typically hexapolar) field, with the ions extracted at one end of the mirror. Performances (for example, the extracted ion current $I$ for a particular charge state) improved dramatically over the last three decades, as summarized by review articles [2, 3, 4]. To a good extent, such improvements are ascribed to the progressively higher microwave frequency ω used for heating, in agreement with the $I\sim\omega^2$ scaling [1]. According to this scaling and to the EC resonant condition, the extracted current improves as $B_{ECR}^2$, where $B_{ECR}$ is the field evaluated at the location where EC heating takes place. In proportion with ω and $B_{ECR}$, the maximum field $B_{max}$ used for axial confinement, evaluated at the ends of the magnetic mirror, was also increased, resulting in increased costs and complexity.

In brief simplistic terms, $B_{ECR}$ sets the performances and $B_{max}$ affects the costs, but the latter is about 3 times higher than $B_{ECR}$, in a conventional linear ECRIS. The reasons for this is axial confinement of particles (apart from a loss-cone in velocity space). By contrast, confinement in a toroidal device does not require such field non-uniformity (in fact, excessive toroidal non-uniformity is actually detrimental to confinement [5]). We therefore propose that a toroidal ECRIS would make a better use of its maximum field $B_{max}$, because $B_{ECR}$ would be just nearly as high as $B_{max}$. Equivalently, a lower $B_{max}$ would be sufficient in a toroidal ECRIS to achieve the same target $B_{ECR}$ as in a linear one.

The toroidal configuration is expected to increase the ion confinement time. The ions that normally would be lost in the longitudinal direction in a linear ECRIS remain confined for a longer time, thus experiencing more collisions and –on average- being ionized to higher charge states. An additional innovative element in our proposed toroidal device is the twisting of the hexapolar coils, which further improves confinement, compared with a simple toroidal device.

The paper is organized as follows. The conceptual design of the toroidal ECRIS is described in Sec.2. Sec.3 enters in more details about the magnetic configuration. Sec.4 presents single-particle calculations of how the ions drift in such configuration, and finally Sec.5 shows how adding a capacitor or a coil could deform the ion trajectories and lead to their extraction

## 2. Toroidal ECRIS design

In its simplest form, a toroidal ECRIS could consist of several linear ECRISs connected in series with each other in a toroidal arrangement (Fig. 1b). This is reminiscent of the ELMO Bumpy Torus fusion experiment [6], consisting of a toroidal array of magnetic mirrors. In analogy with linear ECRIS we could super-impose to the bumpy torus configuration a toroidally curved hexapole field with the goal of improving radial confinement. Such hexapole field could be generated by 6 circular coils, reminiscent of the Poloidal Field (PF) coils in a tokamak, although energized in alternate directions. The PF coils are shown in Fig. 1b, along with several Toroidal Field (TF) coils. An important difference with the tokamak, however, is the absence of toroidal plasma current in this toroidal ECRIS. Therefore, for good confinement we should rather follow the analogy with a toroidal device which has no plasma current, namely the classical stellarator [7], in which the PF coils are now helically twisted (Fig. 1c). This results in the helical twisting of the magnetic field lines, or rotational transform, that helps confinement [7]. Furthermore, we propose replacing the discrete TF coils with a single toroidal coil referred to as "mono-coil", depicted in Fig. 1c and expected to generate no "ripple". In fact, it is proposed that the toroidal vacuum vessel itself can also serve as single TF coil, as in the Madison Symmetric Torus (MST) [8], with added benefits of compactness, simplicity and diagnostic access. To serve all purposes, the vacuum vessel should be made of Aluminium or other highly electrically conductive metal with reasonable stress-strain properties, coated with a plasma-facing material with good resistance to sputtering, such as Chromium.

The metallic vessel should feature a toroidal cut, for instance on the outer equator, filled with an electrically insulating vacuum seal. Biasing the upper outer equator relative to the lower outer equator would result in a current flowing in the vessel in the poloidal direction, generating a magnetic field in the toroidal direction.

To fix the ideas, here we consider a compact tabletop device of major radius 35 cm and minor radius 7.5 cm, with 2 cm thick walls. We examined coil configurations of both types in Fig. 1b and 1c. For the configuration in Fig. 1c, we have considered hexapoles (6 coils energized in alternate directions, or $l=3$, in stellarator jargon) helically wound with toroidal mode number $n=2$, 3 or 4 and quadrupoles ($l=2$) of $n=3$.

## 3. Magnetic configuration

### 3.1 Toroidal field

In the remainder we set the toroidal field to evaluate $B_{tor}=2$ T on the magnetic axis, as this value is intermediate between typical $B_{ECR}$ (~1 T) and $B_{max}$ (~3T) state-of-the-art ECRIS [2-4]. The corresponding frequency for ECR heating at the fundamental harmonic would be a record high 56 GHz. Coil-currents and fields presented hereafter can be easily, linearly rescaled to a different field, $B_{tor} \neq 2$ T, or different frequency, $f \neq 56$ GHz.

A voltage-difference of 0.73 V between the upper and lower outer equators of the vessel causes a poloidal current $I_{pol}= 3.5$ MA to flow through it and generate a field $B_{tor} =2$ T in the plasma center. As expected, and as illustrated in Fig. 2, the field is significantly more uniform in the mono-coil configuration depicted in Fig. 1c than in the discrete TF coil configuration of Fig. 1b. In particular, the toroidal ripple is cancelled at all locations, including outer radii $R$ where it is normally more pronounced, due to proximity to the coils. Furthermore, in the mono-coil device $B_{tor}$ decays like $1/R$ at all toroidal location. Discrete coils, instead, introduce toroidal non-uniformities at large and small $R$ (Fig. 2b).

Particle tracing calculations showed that these reduced toroidal ripples resulted in reduced particle losses, in agreement with tokamak observations.

### 3.2 Hexapolar field

The poloidal field $B_{pol}$ is exerted by a set of 6 coils energized in alternate directions (i.e., the current in a coil is opposite to currents in the two adjacent coils), as in a classical stellarator of $l=3$.

The multipolar field creates a magnetic well that radially confines the plasma. Among several possible multipoles, the majority of modern linear ECRIS settled on hexapoles, as a compromise between ion confinement (favoured in lower order multipoles) and ion extraction (favoured in multipoles of higher order, characterized by a larger area for extraction at the end of the magnetic mirror). Basically the ion confinement needs to be good, for obvious reasons illustrated by the Golovaninski plot [1], but not "too good", because eventually the ions need to be extracted.

For the first toroidal ECRIS we also opted for a hexapole, although helically twisted. The main reason is direct comparison with most linear ECRIS. Considerations on the larger area of longitudinal extraction do not apply to a toroidal ECRIS, where the extraction is radial. Like a linear ECRIS, we also want the ion confinement to be "good but not too good", in the sense meant above. However, this compromise is realized by adopting a classical stellarator configuration, as a compromise between a simple magnetized torus [9] (Fig.1b) and an advanced modular coil stellarator [7]. Once the configuration is chosen, its multipole order $2l$ has a comparatively smaller effect on confinement [7].

Calculations of the magnetic well associated with an $n=3$ hexapole are presented in Fig. 3. The field is up-down symmetric, as expected (Fig. 3c), and is stronger at outer radii than on the inboard side (Fig. 3b). This is due to the difference in inclination of the helical coils on the inboard and outboard side: as evidenced in Fig. 3c, on the inboard side the coils are more vertical, and the poloidal field is reduced as a result.

The field-line map deviates from the typical hexapole in that the center of symmetry is displaced toward inner radii (Fig. 4a). This is due to the toroidal deformation of the hexapole. Furthermore the field pattern rotates as expected, while moving in the toroidal direction, as a consequence of the hexapolar coils being twisted (Fig. 4b).

Magnetic well profiles (Fig. 5e) were generated for various $n=2$, 3 and 4 hexapole and $n=3$ quadrupole configurations (Figs. 5a-d). Apart from the effects illustrated in Figs. 3 and 4, the magnetic wells in Fig. 5 are consistent with hexapole and quadrupole wells in linear ECRIS [1]. Poloidal fields as high as 0.1-1T at the outer plasma edge can be obtained for currents of 30-300 kA in the hexapolar coils.

**3.3 Toroidal and Hexapolar field combined**

Increasing the poloidal field strength leads to an increasingly more defined polygonal shape of the plasma and to a progressively smaller plasma volume (constrained by the axisymmetric vessel), as seen in Fig. 6.

Similar shapes, not shown, were obtained for configurations of different $n$ and $l$. From a top view they all look like polygons of $nl$ sides, but their actual 3D shapes are obviously more complicated.

## 4. Single Particle Tracing

Single particle studies are presented here, of ions only interacting with the applied fields, but not with each other nor with electrons. The knowledge of these single particle trajectories is necessary to set an upper bound to the ion confinement time. They are also a pre-requisite for the design and optimization of ion-extraction techniques. To this end, single-particle-tracing calculations were performed in 3D in a toroidal field (Sec.4.1) as well as in the full geometry under consideration (Sec.4.2) with the aid of the COMSOL Multi-Physics code. This is a finite element code, but it also contains modules solving the equation of motion for a single charged particle. Only the Lorentz force acting on a single ion was considered. Multi-particle effects such as turbulence, collisions or pressure gradients were not included in the present simulations. Ions of different mass, charge and initial location were considered, but they all had the same initial kinetic energy, 2 eV, and all had a very specific pitch angle, such that $v_\perp = \sqrt{2} v_{II}$ where $v_\perp$ and $v_{II}$ are the initial ion velocity components respectively perpendicular and parallel to the magnetic field. Different energies and pitch-angles (or, equivalently, different $v_\perp$ and $v_{II}$) will be considered in the future. Different charge states were already considered but, in the present single-particle treatment, they simply confirmed the well-known fact that charge-dependent drift velocities grow linearly with the charge. Different charge states will have to be reconsidered in the future in a multi-particle framework in order to reproduce the richer physics of highly charged ions in an ECRIS plasma. For example, ions of higher charge are known to benefit from longer confinement times, and are extracted at higher rates in the afterglow after microwave heating is turned off [1]. Multi-particle self-consistent confinement-time estimates for ions of different charge states, combined with electron density and temperature profiles, will also enable predictions of ion density profiles for different charge states [10].

To give an idea of the timescales, the Larmor frequency for an Ar$^+$ ion evaluated on the magnetic axis ($B_{tor}$= 2 T) is about 3 MHz; for the energy and pitch-angle chosen, the ion completes a toroidal orbit in 1.2 ms.

For the same energy and pitch-angle considered, the characteristic Spitzer collision frequencies are about 160 kHz for electron-electron collisions; 60 kHz for electron-ion collisions; 6 MHz for ion-ion collisions. Ion-electron collision are negligibly rare.

## 4.1 Ion motion in a toroidal field

An immediate difference with a linear ECRIS is that ions in a toroidal field are subject to a curvature drift. This is visible in Fig. 7a, in the form of a relatively slow, vertical drift of an $Ar^+$ ion. The ion Larmor radius of about 400 µm can also be recognized in the figure.

Fig. 7d shows a puncture plot on the $\varphi= 0°$ cross section for $Ar^+$ ion trajectories in different initial positions inside the plasma chamber. Small horizontal deviations from the otherwise vertical trajectory, as seen in the puncture plot, are due to the finite Larmor radius.

The calculated curvature drift velocity, about 3.8 m/s, is in very good agreement with the simulations of Fig.7a and d. This drift velocity is about the same for $Ar^+$ ions with the same velocity but located elsewhere in the toroidal plasma, due to the fact that the product of the magnitude and curvature radius of the magnetic field are the same.

## 4.2 Ion motion in a toroidal and hexapolar field

Adding the hexapolar field bends the trajectories presented in the previous section. The bending is illustrated in Fig.7b and e, and is more pronounced on the outboard side, where the twisted hexapolar field has a stronger vertical component and therefore a stronger effect. By contrast, on the inboard side the twisted helical field is mainly oriented in the toroidal direction, and the configuration resembles the purely toroidal configuration described in the previous section, yielding a purely vertical drift. Interestingly, a sizeable population of ions is interested by such a strong trajectory-bending that those ions are not lost by vertical drift. Instead, they are confined for as long as we could simulate (hundreds of ms) in our idealized single-particle (thus, collisionless) model. Confined ion trajectories design non-axisymmetric "tubes" of vaguely triangular or D-shaped poloidal cross section. This shape rotates and evolves as we move around toroidally. These tubes associated with particle trajectories do not coincide with the flux surfaces designed by field-lines, due to drifts. This change in shape of the ion trajectories with increasing poloidal field strength is in agreement with the observation of a more defined polygonal plasma shape (Fig.6) and to a progressively smaller plasma volume, as the non-axisymmetric plasma is "limited" by the axisymmetric vessel.

At very high hexapole currents the plasma becomes highly three-dimensional, which might have interesting consequences for ion losses and extraction (Fig.7c).

## 5. Beam extraction methods

In a linear ECRIS the loss cone of the magnetic mirror in the longitudinal direction is used to electrostatically collect positive ions by means of a negatively biased electrode. A toroidal ECRIS, on the other hand, is toroidally closed on itself, which improves the confinement of ions, electrons and energy, but makes ion extraction more difficult. Ions in a toroidal ECRIS have to be extracted radially.

### 5.1 Extraction by E × B drift

The application of a poloidal electric field **E** by means of a capacitor placed at the edge of the magnetized plasma results in an **E**x**B** drift, mostly radial, because **B** is mostly toroidal. The resulting flow extracts plasma from the torus, which can then be electrostatically or magnetically separated in electrons and ions of interest.

The extractor is toroidally curved and placed closely to the plasma torus, on its inboard side. In these preliminary simulations it was chosen to have a toroidal extent of 14° in order to fit in the gap between two coil in the $l=2$, $n=3$ case, and the plates were assumed to lie at 2 cm from each other (Fig.8). Ions transiting near the capacitor interact with its fringing field, which is shielded by the plasma (Fig.9) but is finite and yields sufficient **E**x**B** drift for the ions to be successfully extracted, as shown by single particle tracing (Fig.8).

The present single-particle simulations do not include plasma effects yet, therefore they do not include the Debye shielding of the applied electric field inside the plasma. However, the strong reduction of E is mimicked by treating the plasma as a dielectric of high relative dielectric permeability, $\varepsilon_r=10^4$. This approximation (red in Fig.9) remains to be improved with respect to the spatial dependence of $E$: the correct dependence, depicted in green in Fig.10, should consist of a rapid decay at the plasma edge over a Debye length of about 700 μm (assuming an electron density of $10^{17}$ m$^{-3}$ and electron temperature of 1 keV). One of the consequences of the approximation is the underestimate of $E$ outside the plasma and in the outer part of the plasma. Note that for the sake of the present paper this underestimate is actually encouraging: if evidence of ion extraction is found with an underestimated $E$ (Fig.9), ion extraction should be even easier in an actual experiment, or require a smaller electric field.

**5.2 Magnetic deflector**

The cylindrical coil depicted in Fig.10a super-imposes a dipole field (Fig.10b) to the existing magnetic configuration.

The coil deflects the ion trajectories until eventually extracting the ions. This is demonstrated in Figs.10c-d for extraction on the midplane, respectively on the inboard and outboard side. The coil considered here is in copper, has a radius of 0.03 m, length of 0.05 m and thickness of 0.01 m. When placed on the inboard side, it is vertically tilted by 25° in order for the straight extraction line to bypass the other side of the torus.

**5.3 Alternate extraction techniques and location of extractor: discussion and future work**

Future single-particle simulations will assess the feasibility of additional ion extraction techniques not considered here. One such technique is electrostatic extraction by means of a biased electrode, as in linear ECRIS. Unlike the **ExB** method of Sec.5.1, this one would involve a single electrode and a radial, rather than poloidal, electric field. A variant of the magnetic extraction of Sec.5.2 could make use of magnetic "divertors", as in tokamaks, or "island divertors", as in stellarators [7]. Placing a collector close enough to the plasma boundary (closer than the finite ion Larmor radius) is also expected to result in ion collection. Finally, techniques used to transfer particles from one storage ring or accelerator to the next will also be considered. Mass and charge selectivity of all these techniques and the above will be studied.

The optimal location of an ion extractor will also be investigated. The natural choice will be to place the extractor in a location where enhanced ion losses naturally occur anyway. The goal is in part automatically achieved by means of drifts (Fig. 7). For example, increased **ExB** extraction was observed if the extractor was placed at lower $z$, but can be further optimized. For this purpose a toroidal and poloidal map of natural ion losses will be developed for the magnetic configuration considered (Fig. 1c).

Optimal placement of the extractor might also depend on the local curvature of the magnetic field and, as a result, of the plasma: in a linear ECRIS these are known to affect the beam shape and emittance; a concave plasma "meniscus" is usually optimal. In a typical axisymmetric toroidal plasma, however, the curvature can be, at best, concave in one direction (toroidal) and convex in the other (poloidal). This convex-concave curvature is obtained on the inboard side of the torus. Indeed, initial evidence suggests extraction to be more efficient on the inboard (Fig.

10c) than on the outboard (Fig. 10d), but more studies are needed and more positions need to be compared. Note that in a highly non-axisymmetric torus or in an axisymmetric torus of special "bean-shaped" cross-section it is possible for the meniscus to be concave in all directions.

**Conclusions**

Electron Cyclotron Resonant Ion Sources (ECRIS) have traditionally been linear devices. A toroidal source was proposed here for the first time, in order to improve the ion confinement and reach higher charge states. At the same time the device makes a more efficient and economical use of the magnetic field. This is expected to facilitate the race to higher fields and higher ECR frequencies, which are connected with more intense ion currents. Single-particle modelling presented here suggests ion extraction to be feasible by at least two techniques, involving respectively a radial **E**x**B** flow and magnetic deflection.

**Acknowledgements**

S. Gammino and C. Lyneis are thankfully acknowledged for the fruitful discussions, as is L. Neri for the suggestion to consider inboard extraction.

**References**

[1] R. Geller, *Electron Cyclotron Resonance Ion Sources and ECR Plasmas*, Institute of Physics Pub, 1996.

[2] D. Hitz, G. Melin, and A. Girard, *Rev. Scientific Instruments* **71**, 839 (2000)

[3] A. G. Drentje, *Rev. Scientific Instruments* **74**, 2631 (2003)

[4] C. Lyneis, D. Leitner, M. Leitner, C. Taylor and S. Abbott, *Rev. Scientific Instruments* **81**, 02A201 (2010)

[5] P.N. Yushmanov, "Diffusive transport processes caused by ripple in tokamaks" in *Reviews of plasma physics* (ed. B.B. Kadomtsev) Vol. 16, p.117, Consultants Bureau, New York (1990)

[6] S. Hiroe *et al.*, *Nucl. Fusion* **28**, 2249 (1988)

[7] M. Wakatani, *Stellarator and Heliotron Devices*, Oxford University Press, 1998

[8] S. C. Prager *et al., Phys. Fluids B* **2**, 1367 (1990)

[9] A. Fasoli *et al.*, *Plasma Phys. Controll. Fusion* **52**, 124020 (2010)

[10] H.R. Griem, *Principles of Plasma Spectroscopy*, Cambridge University Press, Cambridge 1997

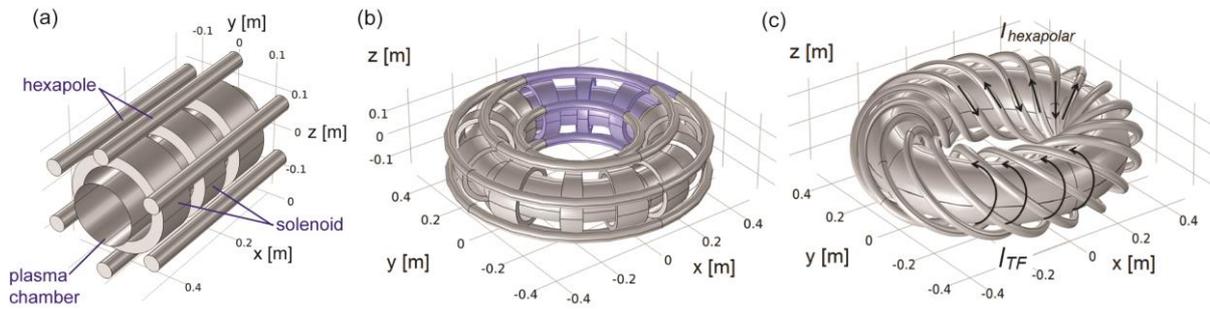

**Figure 1.** (a) Schematic view of a linear Electron Cyclotron Resonance Ion Source (ECRIS). (b) Toroidal assembly of several ECRISs connected in series with each other. (c) Similar to (b), except that discrete Toroidal Field (TF) coils are replaced by currents flowing in the poloidal direction in the electrically conductive vacuum vessel, and hexapole coils are helically deformed to generate rotational transform and improve confinement.

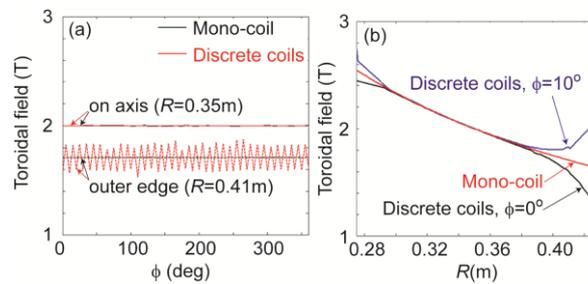

**Figure 2.** Illustration of (a) toroidal non-uniformity and (b) radial profile of toroidal field in a device featuring discrete TF coils (Fig.1b) or whose vacuum vessel acts as a "mono-coil" (Fig.1c).

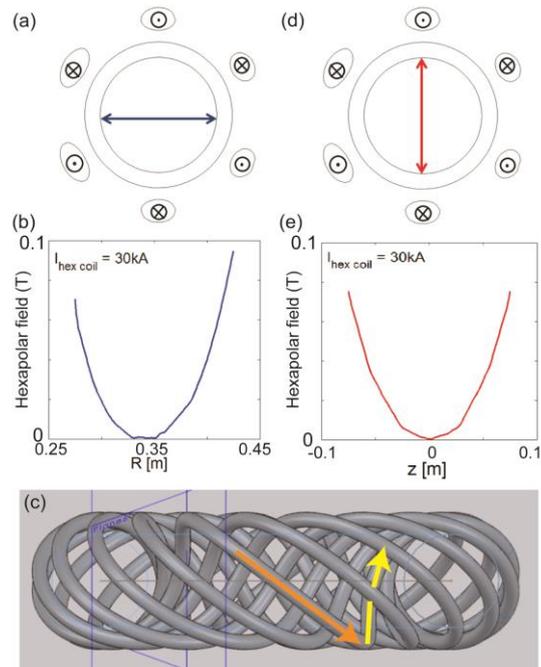

**Figure 3.** (a) Poloidal cross-section of vessel and hexapole coils, with diameter marked. (b) Profile of hexapole field along marked diameter. (c) Side-view explains why $B_{hex}$ is stronger at outer radii: coils are more horizontal on outboard side (orange), thus generating stronger poloidal field. (d) Like (a), but with vertical marked. (e) Like (b), but along vertical.

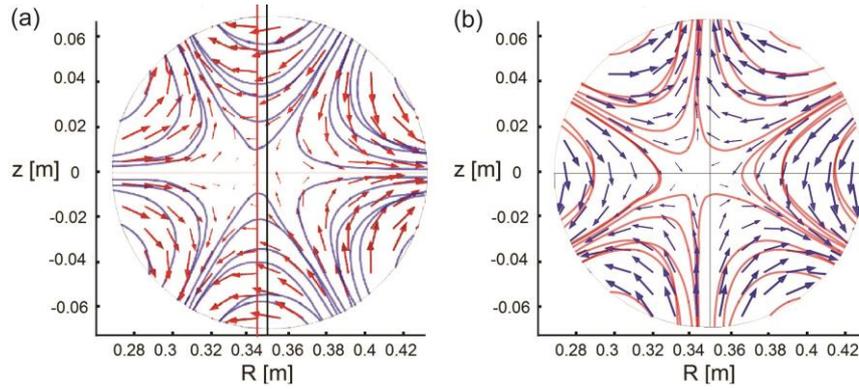

**Figure 4.** (a-b) Visualization of hexapole field in poloidal cross-sections at $\varphi = 0°$ and $\varphi = 30°$. Note that the center of the cusp is displaced relative to $R=0.35$ m due to toroidicity.

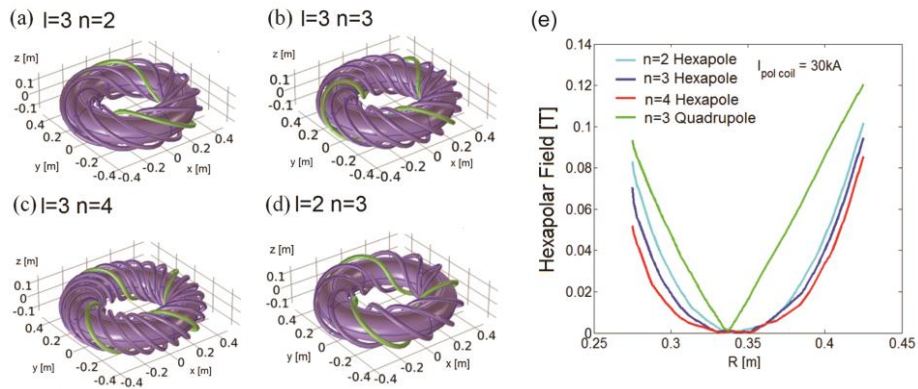

**Figure 5.** (a-d) Bird's-eye views of configurations of different $l$ and $n$ and (e) radial profiles of corresponding "magnetic wells".

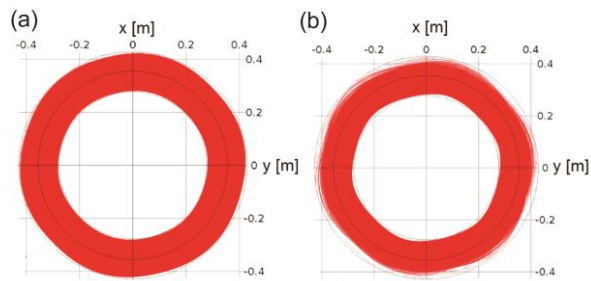

**Figure 6.** Top view of configuration in Fig.5a ($l=3$, $n=2$), for hexapolar coil current of (a) 30 kA and (b) 300 kA.

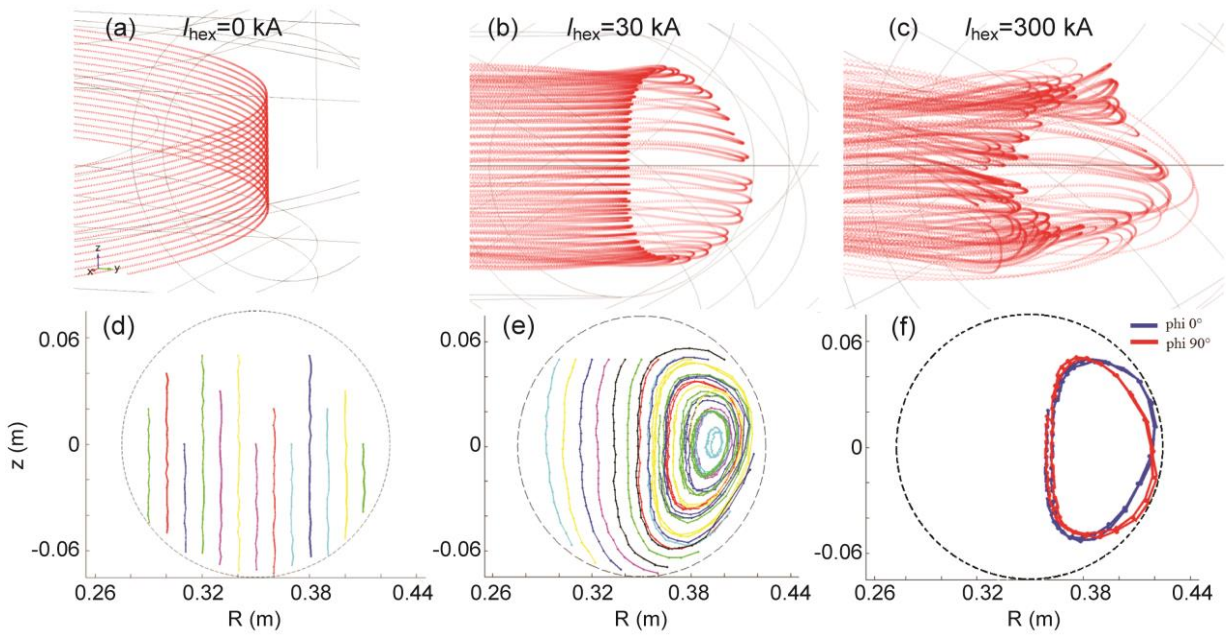

**Figure 7**. (a-c) Computed Ar+ orbits for zero, moderate and high hexapolar field. (d-e) Puncture plots, at $\varphi = 30º$, of the Ar+ ion trajectories of Figs.(a-b). The trajectories do not look smooth because the Larmor radius is finite (700 μm). (f) Detail of the closed D-shaped tube formed by ion trajectories, at different toroidal locations $\varphi$.

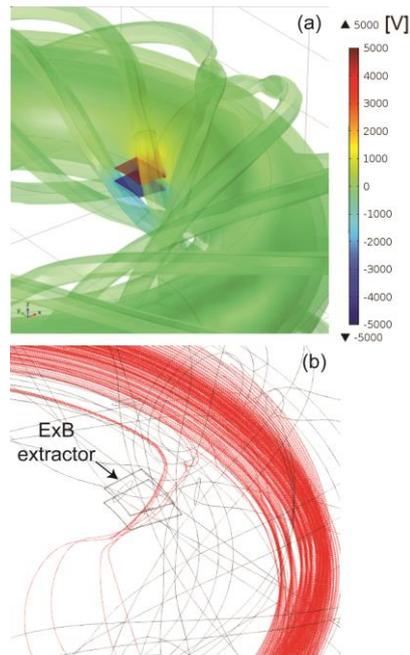

**Figure 8.** (a) Schematic of two oppositely biased plates, applying a vertical electric field **E** at the plasma edge, on its inboard side. (b) Demonstration that ions transiting close to capacitor are extracted by **E**x**B** drift.

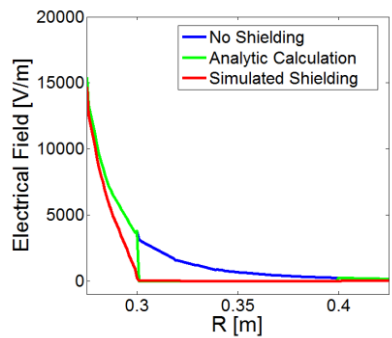

**Figure 9.** Radial profiles of electric field generated by electrodes located at R<0.3m and biased at ±1kV in presence (green) or absence (blue) of the plasma, which has a shielding effect. Plasma effects are not included in these preliminary single-particle studies, but the ad hoc introduction of a dielectric mimics well the electric field inside the plasma, at *R*>0.3m.

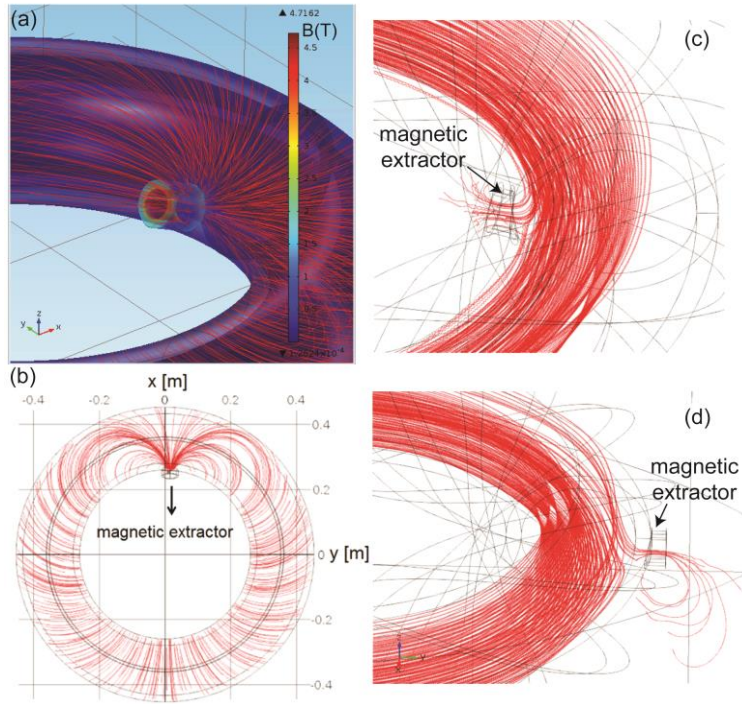

**Figure 10.** (a-b) bird's-eye view and equatorial cross-section of field-lines generated by a circular coil extractor placed on inboard side. (c-d) Computed ion trajectories showing extraction on inboard and outboard side.